\documentclass{elsart}                                               
\usepackage{amssymb}
\journal{Nuclear Physics B}
\begin{document}
\begin{frontmatter}
\title{Superconformal
Field Theory and SUSY $N=1$ KdV Hierarchy II:
The Q-operator}
\author[POMI]{Petr P. Kulish}
\ead{kulish@pdmi.ras.ru}
\author[SpbU]{Anton M. Zeitlin\corauthref{cor1}}
\ead{zam@math.ipme.ru} 
\ead[url]{http://www.ipme.ru/zam.html}
\address[POMI]{St. Petersburg Department of Steklov Mathematical 
Institute, Fontanka, 27,\\ St. Petersburg, 191023, Russia}
\address[SpbU]
  {Department of High Energy Physics, Physics Faculty, St. Petersburg State 
  University, Ul'yanovskaja 1, Petrodvoretz, St.Petersburg, 198904, Russia}
\corauth[cor1]{Corresponding author}
\date{}
\begin{abstract}
The algebraic structures related with integrable structure of superconformal field theory (SCFT) are introduced. The SCFT counterparts of Baxter's Q-operator are constructed. The fusion-like relations for the transfer-matrices in different representations and their truncations are obtained.   
\end{abstract}

\end{frontmatter}
\section{Introduction}
 During last decade the integrable structures associated with conformal field theory (CFT) and its $W_3$ extension were studied \cite{1}-\cite{5}. The main problem to which these papers were addressed is the simultaneous diagonalization of the corresponding infinite family of the integrals of motion (IM).
 The method used for this purpose is the continuous version of the Quantum Inverse Scattering Method (QISM) \cite{leshouches}. 
The most successful results were obtained by 
means of the generalization of the Baxter's 
$\mathbf{Q}$-operator relation \cite{baxter}.
 In this paper we will follow this way constructing the $\mathbf{Q}$-operator for analysis of the integrable structure arising in the superconformal field theory (SCFT). This CFT includes both usual left and right components of the energy-momentum tensor $T(u)$, $\bar{T}(\bar{u})$ and spin-$\frac{3}{2}$ local fields $G(u)$, 
$\bar{G}(\bar{u})$ where variable $u$ 
lies on a cylinder of circumference $2\pi$.
The  full symmetry algebra of the theory is $SCA\times\overline{SCA}$ ($SCA$ means superconformal algebra).We will concentrate here on the left-chiral component-
SCA. The boundary conditions are: 
$$
T(u+2\pi)=T(u),\quad G(u+2\pi)=\pm G(u),
$$
where ``+'' corresponds to Ramond (R) sector and ``--'' to Neveu-Schwarz (NS) one. The $SCA$ gives possibility to construct two infinite abelian families of conserved charges \cite{bilal},\cite{plb1},\cite{plb2} where one is SUSY invariant and the other is not.
Here we will consider the integrable structure invariant under the supersymmetry transformation. In the classical limit the corresponding IM give the Hamiltonians of the SUSY $N$=1 KdV hierarchy \cite{inami} based on twisted affine 
Lie superalgebra 
$sl(2|1)^{(2)}\cong osp(2|2)^{(2)}\cong C(2)^{(2)}$ \cite{kac}. 
The supertrace of the monodromy matrix of the associated L-operator in the 3-dimensional representation of $C(2)^{(2)}$ (see Appendix 1) is a generating function for the IM. Thus it's quantum generalization, the transfer matrix produces the IM of the SCFT.  
 We will review some results from part I \cite{plb2} concerning the construction of the quantum counterpart of the monodromy matrix and it's basic properties in Sec. 2. Using the super q-oscillator representations of the upper Borel subalgebra of quantum affine superalgebra $C_{q}(2)^{(2)}$ we define the $\mathbf{Q}_{\pm}$ operators. The transfer matrices in different evaluation representations can be expressed in such a way:
$$
2cos(\pi P)\mathbf{t}_s(\lambda)=\mathbf{Q}_{+}(q^{s+\frac{1}{4}}\lambda)\mathbf{Q}_{-}(q^{-s-\frac{1}{4}}\lambda)+\mathbf{Q}_{+}(q^{-s-\frac{1}{4}}\lambda)\mathbf{Q}_{-}(q^{s+\frac{1}{4}}\lambda),
$$
where $s$ runs over integer and half-integer nonnegative numbers. $\mathbf{Q}_{\pm}$ operators satisfy quantum super-Wronskian relation:
$$
2cos(\pi P)=\mathbf{Q}_{+}(q^{\frac{1}{4}}\lambda)\mathbf{Q}_{-}(q^{-\frac{1}{4}}\lambda)+\mathbf{Q}_{+}(q^{-\frac{1}{4}}\lambda)\mathbf{Q}_{-}(q^{\frac{1}{4}}\lambda).
$$ 
One should note, that we use only $4s+1$-dimensional 
``$osp(1|2)$-induced'' representations (sometimes called atypical) of $C(2)^{(2)}$. It allows, however, to construct the fusion relations, see below.
To construct the Baxter-like relation we introduce additional ``quarter''-operators, constructed ``by hands'' from the $\mathbf{Q}$-operators:
$$
2cos(\pi P)\mathbf{t}_{\frac{k}{4}}(\lambda)=\mathbf{Q}_{+}(q^{\frac{k}{4}+\frac{1}{4}}\lambda)\mathbf{Q}_{-}(q^{-\frac{k}{4}-\frac{1}{4}}\lambda)-
\mathbf{Q}_{+}(q^{-\frac{k}{4}-\frac{1}{4}}\lambda)
\mathbf{Q}_{-}(q^{\frac{k}{4}+\frac{1}{4}}\lambda)
$$
for odd integer $k$.
 The Baxter's relations are:
$$
\mathbf{t}_{\frac{1}{4}}(\lambda)\mathbf{Q}_{\pm}(\lambda)=\pm \mathbf{Q}_{\pm}(q^{\frac{1}{2}}\lambda)\mp\mathbf{Q}_{\pm}(q^{-\frac{1}{2}}\lambda),
$$
$$
\mathbf{t}_{\frac{1}{2}}(q^{\frac{1}{4}}\lambda)\mathbf{Q}_{\pm}(\lambda)=\mathbf{t}_{\frac{1}{4}}(q^{\frac{1}{2}}\lambda)\mathbf{Q}_{\pm}(q^{-\frac{1}{2}}\lambda)+\mathbf{Q}_{\pm}(q\lambda).
$$
The fusion relations have the following form very similar to the $A_1^{(1)}$ case \cite{1}:
$$
\mathbf{t}_{j}(q^{\frac{1}{4}}\lambda)\mathbf{t}_{j}(q^{-\frac{1}{4}}\lambda)=
\mathbf{t}_{j+\frac{1}{4}}(\lambda)\mathbf{t}_{j-\frac{1}{4}}(\lambda)+(-1)^{4j}.  
$$
But these relations are only ``fusion-like'' because the ``quarter''-operators do not seem to correspond to any representation of $C(2)^{(2)}$. This will be the subject of Sec. 3. Finally in sec. 4 we describe the truncation of these relations for different values of $q$, being root of unity.
\section{ Quantum Monodromy Matrix and its Properties}
In the part I \cite{plb2} we have explicitly constructed the quantum counterpart of the monodromy matrix of the linear problem for the L-operator of the SUSY N=1 KdV equation. This L-operator has the following explicit form:
\begin{eqnarray}\label{lop}
\mathcal{L}_F=D_{u,\theta} 
-D_{u,\theta}\Phi h_{\alpha}-(e_{{\delta-\alpha}}+ e_{{\alpha}}),
\end{eqnarray}
where $h_{\alpha}$, $e_{{\delta-\alpha}}\equiv e_{\alpha_0}$, 
$e_{{\alpha}}$ are the 
Chevalley generators of $C(2)^{(2)}$ (see Appendix 1), 
$D_{u,\theta} =\partial_\theta + \theta \partial_u$ 
is a superderivative, the variable 
$u$ lies on a cylinder of circumference $2\pi$, $\theta$ 
is  a  Grassmann  variable, $\Phi(u,\theta)=\phi(u) - 
\frac{i} {\sqrt{2}}\theta\xi(u)$ 
is a bosonic superfield.
The Poisson brackets for the field $\Phi$, obtained by means of the 
Drinfeld-Sokolov
reduction are:
\begin{equation}
\{D_{u,\theta}\Phi(u,\theta), D_{u',\theta'}\Phi(u',\theta')\}=  
D_{u,\theta}(\delta(u-u')(\theta-\theta'))
\end{equation}
and the following boundary conditions are imposed on the components of 
$\Phi$: $\phi(u+2\pi)=\phi(u)+2\pi i p$, $\xi(u+2\pi)=\pm\xi(u)$.
Making a gauge transformation of the auxiliary linear problem 
we obtain a new superfield $\mathcal{U}(u,\theta)\equiv
D_{u,\theta}\Phi(u,\theta)\partial_u\Phi(u,\theta)-D_{u,\theta}^3
\Phi(u,\theta)=-\theta U(u)-i\alpha(u)/\sqrt{2}$, 
where $U$ and $\alpha$ generate the superconformal algebra under the Poisson 
brackets:
\begin{eqnarray}
\{U(u),U(v)\}&=&
 \delta'''(u-v)+2U'(u)\delta(u-v)+4U(u)\delta'(u-v),\\
\{U(u),\alpha(v)\}&=&
 3\alpha(u)\delta'(u-v) + \alpha'(u)\delta(u-v),\nonumber\\
\{\alpha(u),\alpha(v)\}&=&
 2\delta''(u-v)+2U(u)\delta(u-v)\nonumber.
\end{eqnarray} 
The corresponding infinite series of the integrals of motion \cite{bilal}: 
\begin{eqnarray}\label{IM}
I^{(cl)}_1&=&\frac{1}{2\pi}\int U(u)\d u,\\
I^{(cl)}_3&=&\frac{1}{2\pi}\int
\Big(U^2(u)+\alpha(u)\alpha'(u)/2\Big)\d u,\nonumber\\
I^{(cl)}_5&=&\frac{1}{2\pi}\int
\Big(U^3(u)-(U')^2(u)/2-\alpha'(u)\alpha''(u)/4-
\alpha'(u)\alpha(u)U(u)\Big)\d u,\nonumber\\
& &   .\qquad.\qquad.\nonumber
\end{eqnarray}
allows us to construct integrable evolution equations. 
For example $I_2$ yields 
the SUSY N=1 KdV equation \cite{mathieu}:
\begin{equation}
\mathcal{U}_t=-\mathcal{U}_{uuu}+3(\mathcal{U} D_{u,\theta}\mathcal{U})_u
\end{equation}
which has the following form in components:
\begin{eqnarray}
U_t&=&-U_{uuu}-6UU_u - \frac{3}{2}\alpha\alpha_{uu}\\ 
\alpha_t&=&-4\alpha_{uuu}-3(U\alpha)_u.\nonumber
\end{eqnarray} 
These integrals of motion (\ref{IM}) can be extracted 
from the monodromy matrix of the
L-operator (\ref{lop})\cite{plb2}:
\begin{eqnarray}
\mathbf{M}^{(cl)}&=&e^{2\pi i ph_{\alpha_1}}
P\exp\int_0^{2\pi} \d u\Big(\frac{i}{\sqrt{2}}
\xi(u)e^{-\phi(u)}e_{\alpha_1}\\
&-&\frac{i}{\sqrt{2}}
\xi(u)e^{\phi(u)}e_{\alpha_0}
-e^2_{\alpha_1}e^{-2\phi(u)}-
e^2_{\alpha_0}e^{2\phi(u)}-[e_{\alpha_1},e_{\alpha_0}]
\Big).\nonumber
\end{eqnarray}
It's quantum generalization can be represented in the quantum P-exponent form 
(for the explanation of this notion see below):
\begin{equation}\label{monodromy}
\mathbf{M}=e^{2\pi iPh_{\alpha_1}}Pexp^{(q)}\int^{2\pi}_{0}\d u
(W_{-}(u)e_{\alpha_1} +W_{+}(u)e_{\alpha_0}).
\end{equation}
Vertex operators $W_{\pm}$ have the following form $W_{\pm}(u)=\int d \theta:e^{\pm\Phi(u,\theta)}:= \mp \frac{i}{\sqrt{2}}\xi(u):e^{\pm\phi(u)}:$, where:
\begin{eqnarray}\label{freefields}
&&\Phi(u,\theta)=\phi(u)-\frac{i}{\sqrt{2}}\theta\xi(u)\\ 
&&\phi(u)=iQ+iPu+\sum_n\frac{a_{-n}}{n}e^{inu},\qquad
\xi(u)=i^{-1/2}\sum_n\xi_ne^{-inu},\nonumber\\
&&[Q,P]=\frac{i}{2}\beta^2 ,\quad 
[a_n,a_m]=\frac{\beta^2}{2}n\delta_{n+m,0},\qquad
\{\xi_n,\xi_m\}=\beta^2\delta_{n+m,0}.\nonumber\\
&&:e^{\pm\phi(u)}:=
\exp\Big(\pm\sum_{n=1}^{\infty}\frac{a_{-n}}{n}e^{inu}\Big)
\exp\Big(\pm i(Q+Pu)\Big)\exp\Big(\mp\sum_{n=1}^{\infty}\frac{a_{n}}{n}e^{-inu}
\Big)\nonumber.
\end{eqnarray} 
$W_{\pm}$ satisfy the following properties:
\begin{eqnarray}
&&W_{\sigma_1}(u_1)W_{\sigma_2}(u_2)=-q^{\sigma_1\sigma_2}W_{\sigma_2}(u_2)W_{\sigma_1}(u_1),\quad q=e^{\frac{\pi i \beta^2}{2}}, \quad u_1>u_2\\
&&PW_{\pm}(u)=W_{\pm}(u)(P\pm\frac{\beta^2}{2})\nonumber
\end{eqnarray}
These relations gave us possibility to show that the reduced universal 
R-matrix of the quantum affine superalgebra $C_q(2)^{(2)}$ (see Appendix 1) 
with lower Borel subalgebra represented by operators 
$(q^{-1}-q)^{-1}\int_{u_2}^{u_1}\d u W_{\pm}(u)$ have the P-exponent multiplication property \cite{plb2}. Thus we have called it the quantum P-exponent:
\begin{eqnarray}
\mathbf{\bar{L}}^{(q)}(u_1,u_2)=Pexp^{(q)}\int^{u_1}_{u_2}\d u
(W_{-}(u)e_{\alpha_1} +W_{+}(u)e_{\alpha_0})\\
\mathbf{\bar{L}}^{(q)}(u_1,u_3)=\mathbf{\bar{L}}^{(q)}(u_1,u_2)
\mathbf{\bar{L}}^{(q)}(u_2,u_3), \quad u_1>u_2>u_3.\nonumber
\end{eqnarray}
The universal R-matrix represented by $(q^{-1}-q)^{-1}\int_{u_2}^{u_1}\d u W_{\pm}(u)$ coincide with
$\mathbf{L}^{(q)}(u_1,u_2)=e^{\pi i P h_{\alpha_{1}}}\mathbf{\bar{L}}^{(q)}(u_1,u_2)$.
Note, that the monodromy matrix is: 
\begin{eqnarray}
\mathbf{M}=e^{\pi i P h_{\alpha_1}}
\mathbf{L}^{(q)}(2\pi,0)
\end{eqnarray}
Unlike the cases of quantum (super)algebra $(A_1^{(1)})_q$, $(A_2^{(2)})_q$, $(A_2^{(1)})_q$, 
$B(0,1)_{q}^{(1)}$ 
\cite{1}-\cite{5}, \cite{plb1} this object can't be 
represented in the usual P-ordered form due to the singularities arising from the products of fermion free fields. However as in all these cases $\mathbf{L}^{(q)}(2\pi,0)$ which we will 
 denote in the following as $\mathbf{L}$ to simplify the notation satisfy the RTT-relation \cite{leshouches}, \cite{kulsklyan}:
\begin{eqnarray}
\mathbf{R}_{ss'}
\Big(\mathbf{L}_s\otimes \mathbf{I}\Big)\Big(\mathbf{I}
\otimes \mathbf{L}_{s'}\Big)
=(\mathbf{I}\otimes \mathbf{L}_{s'}\Big)
\Big(\mathbf{L}_s\otimes \mathbf{I}\Big)\mathbf{R}_{ss'},
\end{eqnarray}
where $s$, $s'$  means that the corresponding object is considered 
in some representation 
of $C_q(2)^{(2)}$. 
Thus the supertraces of monodromy matrix (``transfer matrices'') 
$\mathbf{t}_s=str\mathbf{M}_s$
 commute. It is very useful to consider the evaluation representations of $C_q(2)^{(2)}$, $\rho_s(\lambda)$, where now symbol $s$ means integer and 
half-integer 
numbers (see Appendix 1). Denoting $\rho_s(\lambda)(\mathbf{M})$ as $\mathbf{M}_s(\lambda)$ 
we find that $\mathbf{t}_s(\lambda)=str\mathbf{M}_s(\lambda)$ commute:
\begin{eqnarray} 
[\mathbf{t}_s(\lambda),\mathbf{t}_{s'}(\mu)]=0
\end{eqnarray}
The expansion of $\log(\mathbf{t}_{\frac{1}{2}}(\lambda))$ in $\lambda$ (the transfer matrix in the fundamental 3-dimensional representation) is believed to 
give us as coefficients the local IM, the quantum counterparts of (\ref{IM}):
\begin{eqnarray}
&&I_{2k-1}=\int_{0}^{2\pi}\d u T_{2k}(u), \quad k=1,2,...\\
&&T_2(u)=T(u), \quad T_4(u)=:T^{2}(u):+\frac{i}{4}:G(u)G'(u):, ...\nonumber
\end{eqnarray}
where the densities $T_{2k}(u)$ are differential polynomials of $T(u)$ and $G(u)$ such that $T_{2k}(u)=:T^{k}(u):+$ terms with derivatives. 
$T(u)$ and $G(u)$ have the following representation in terms of free fields 
(\ref{freefields}) (see \cite{SCFT}):
\begin{eqnarray}
&&-\beta^2T(u)=:\phi'^2(u):+(1-\beta^2/2)\phi''(u)+\frac{1}{2}:\xi\xi'(u):+\frac{\epsilon\beta^2}{16}\\ 
&&\frac{i^{1/2}\beta^2}{\sqrt{2}}G(u)=\phi '\xi(u)+(1-\beta^2/2)\xi '(u),
\nonumber
\end{eqnarray}
One of the basic features of the IM is that their commutators with certain vertex operators reduce to total derivatives \cite{feigin}:
\begin{eqnarray}
[I_{2k-1}, W_{\pm}(u)]=\partial_u(:O^{(k)}_{\pm}(u) 
W_{\pm}(u):)=\partial_u\Theta^{(k)}_{\pm}(u)
\end{eqnarray}
(where $O^{(k)}_{\pm}(u)$ is the polynomial of $\partial_u\phi(u)$, $\xi(u)$ and their derivatives). By means of this important property one can show (see \cite{5} ) that such integrals commute with the transfer matrices:
\begin{eqnarray}
[I_{2k-1}, \mathbf{t}_s(\lambda)]=0.
\end{eqnarray}
In our case it is also important to show that the generator of the supersymmetry transformation
$G_0=\int_0^{2\pi}G(u)du$ commutes with $\mathbf{t}_s(\lambda)$.
Really, 
\begin{eqnarray}
[G_0,W_{\pm}(u)]=\pm i^{1/2}\partial_u :e^{\pm\phi(u)}:=\partial_u\Xi^{\pm}(u). \end{eqnarray}
Thus, $G_0$ can be included in the family of IM and using again the results of 
\cite{5} it is easy to show that 
\begin{eqnarray}
[G_0,\mathbf{t}_s(\lambda)]=0.  
\end{eqnarray}
\section{The Construction of the Q-operator and the Fusion Relations}
Let's consider evaluation representations $\rho_s(\lambda)$ of the quantum algebra $C_q(2)^{(2)}$ which we have already mentioned in the introduction and in previous section. They are $4s+1$ dimensional where $s$ runs over integer and half-integer numbers. We call these representations ``$osp_q(1|2)$-induced'' because each triple $\{h_{\alpha_1},e_{\alpha_1},e_{-\alpha_1}\}$ and $\{h_{\alpha_0},e_{\alpha_0},e_{-\alpha_0}\}$ form $osp_q(1|2)$ superalgebra \cite{kulresh},
and $4s+1$-dimensional representation 
space is a representation space for each $osp_q(1|2)$ subalgebra irreducible 
representation (see Appendix 1).\\
\hspace*{5mm}We will also use in the following the corresponding reducible 
infinite-dimen\-sional representations 
$\rho^{+}_s(\lambda)$, which look like $osp_q(1|2)$ Verma representations 
for the $\{h_{\alpha_1},e_{\alpha_1},e_{-\alpha_1}\}$ triple.
To simplify the notation we denote $\rho_s(\lambda)(\mathbf{L})$ as $\mathbf{L}_s(\lambda)$ and $\rho^{+}_s(\lambda)(\mathbf{L})$ as $\mathbf{L}^{+}_s(\lambda)$.\\
\hspace*{5mm}In order to construct the $\mathbf{Q}$-operator we introduce the auxiliary object, the super q-oscillator algebra \cite{q-osc} with the following commutation relations:
\begin{equation}\label{q-osc}
[H,\varepsilon_{\pm}]=\pm\varepsilon_{\pm}, \qquad q^{\frac{1}{2}}\varepsilon_{+} \varepsilon_{-}+q^{-\frac{1}{2}}\varepsilon_{-}\varepsilon_{+}=\frac{1}{q-q^{-1}}. 
\end{equation}
Using these relations it is possible to build realization for the upper Borel subalgebra $b_{+}$ generated by $\{h_{\alpha_1},h_{\alpha_0},e_{\alpha_1},e_{\alpha_0}\}$:
\begin{equation}\label{chi}
\chi_{\pm}(\lambda):\quad e_{\alpha_1}\to\lambda\varepsilon_{\pm},\quad e_{\alpha_0}\to\mp\lambda\varepsilon_{\mp},\quad h_{\alpha_1}=-h_{\alpha_0}\to \pm H
\end{equation}
and we denote $\chi_{\pm}(\lambda)(\mathbf{L})$ as $\mathbf{L}_{\pm}(\lambda)$.
Next, following \cite{2} we define two operators which do not depend on representation of oscillator algebra but depend on (and completely determined by) the commutation relations and cyclic property of the supertrace:
\begin{equation} 
\mathbf{A}_{\pm}(\lambda)=Z^{-1}_{\pm}(P)str_{\chi_{\pm}(\lambda)}(e^{\pm\pi i PH}\mathbf{L}_{\pm}(\lambda)),
\end{equation}
where
\begin{equation}
Z^{-1}_{\pm}(P)=str_{\chi_{\pm}(\lambda)}(e^{\pm 2\pi i PH})
\end{equation}
Multiplying two $\mathbf{A}$ operators we arrive to such a result (see Appendix 2):
\begin{equation}\label{AA}
\mathbf{A}_{+}(q^{s+\frac{1}{4}}\lambda)\mathbf{A}_{-}(q^{-s-\frac{1}{4}}\lambda)=2cos(\pi P)e^{-4\pi i P(s+\frac{1}{4})}\mathbf{t}^{+}_s(\lambda),
\end{equation}
where $\mathbf{t}^{+}_s(\lambda)$ is the supertrace of the monodromy matrix in 
$\rho^{+}_s(\lambda)$ representation.
Defining then the $\mathbf{Q}$-operators:
\begin{equation}
\mathbf{Q}_{\pm}(\lambda)=\lambda^{\pm\frac{2P}{\beta^2}}
\mathbf{A}_{\pm}(\lambda)
\end{equation}
we find that the expression (\ref{AA}) can be written in such a form:
\begin{equation}
\mathbf{Q}_{+}(q^{s+\frac{1}{4}}\lambda)\mathbf{Q}_{-}(q^{-s-\frac{1}{4}}\lambda)=2cos(\pi P)\mathbf{t}^{+}_s(\lambda).
\end{equation}
Now let's recall that $\mathbf{t}^{+}_s(\lambda)$ corresponds to infinite-dimensional representation $\rho^{+}_s$. In order to obtain the transfer matrix corresponding to the finite-dimensional representation $\rho_s$ one should 
consider factor-representation\\
\noindent $\rho^{+}_s$/$\rho^{+}_{-s-\frac{1}{2}}\cong
\rho_s)$ (see Appendix 1). Thus one obtains:
\begin{equation}\label{ttt}
\mathbf{t}^{+}_s(\lambda)+\mathbf{t}^{+}_{-s-\frac{1}{2}}(\lambda)=
\mathbf{t}_s(\lambda).
\end{equation}
Sign + in this formula appears because of taking a supertrace. Really, 
$\rho^{+}_{-s-\frac{1}{2}}$ subrepresentation in $\rho^{+}_s$ representation 
space has a highest weight vector which is odd (though our convention 
is that highest 
weight vector in $\rho^{+}_j$ representation 
is even (see Appendix 1)). That's why
we have sign + instead of the usual --.\\   
\hspace*{5mm}Rewriting (\ref{ttt}) in terms of the $\mathbf{Q}_{\pm}$ operators we find:
\begin{equation}\label{tQ}
2cos(\pi P)\mathbf{t}_s(\lambda)=\mathbf{Q}_{+}(q^{s+\frac{1}{4}}\lambda)\mathbf{Q}_{-}(q^{-s-\frac{1}{4}}\lambda)+\mathbf{Q}_{+}(q^{-s-\frac{1}{4}}\lambda)\mathbf{Q}_{-}(q^{s+\frac{1}{4}}\lambda).
\end{equation}
Considering the case of 1-dimensional representation of $C_q(2)^{(2)}$ ($s=0$)
we obtain a quantum super-Wronskian (qsW) relation:
\begin{equation}\label{sW}
2cos(\pi P)=\mathbf{Q}_{+}(q^{\frac{1}{4}}\lambda)\mathbf{Q}_{-}(q^{-\frac{1}{4}}\lambda)+\mathbf{Q}_{+}(q^{-\frac{1}{4}}\lambda)\mathbf{Q}_{-}(q^{\frac{1}{4}}\lambda).
\end{equation}
In order to 
construct the Baxter type relation \cite{baxter} we introduce the auxiliary
``quarter'' operators:
\begin{eqnarray}\label{quarter}
&&2cos(\pi P)\mathbf{t}_{\frac{k}{4}}(\lambda)=\\
&&\mathbf{Q}_{+}(q^{\frac{k}{4}+\frac{1}{4}}\lambda)\mathbf{Q}_{-}(q^{-\frac{k}{4}-\frac{1}{4}}\lambda)-
\mathbf{Q}_{+}(q^{-\frac{k}{4}-\frac{1}{4}}\lambda)
\mathbf{Q}_{-}(q^{\frac{k}{4}+\frac{1}{4}}\lambda)\nonumber
\end{eqnarray}
for odd integer $k$.
Let's consider the $\mathbf{t}_{\frac{1}{4}}(\lambda)$ operator and multiply
it on $\mathbf{Q}_{\pm}(\lambda)$ operators. Using (\ref{sW}) one can get
the following relation of Baxter type, which is similar to that one obtained in $A_1^{(1)}$ case \cite{2}:
\begin{equation}\label{baxter1}
\mathbf{t}_{\frac{1}{4}}(\lambda)\mathbf{Q}_{\pm}(\lambda)=\pm \mathbf{Q}_{\pm}(q^{\frac{1}{2}}\lambda)\mp\mathbf{Q}_{\pm}(q^{-\frac{1}{2}}\lambda).
\end{equation}
The main difference is that unlike the $A_1^{(1)}$ case we are not seeking 
for the eigenvalues of $\mathbf{t}_{\frac{1}{4}}(\lambda)$ operator but we are interested in the operator corresponding to the 3-dimensional representation, $\mathbf{t}_{\frac{1}{2}}(\lambda)$, because we know that it is a generating function of  
the local IM \cite{plb2}. Thus we are looking for the relation of Baxter type which includes this operator.\\
\hspace*{5mm}In order to do this we need to proceed through some additional 
steps.
First, let's unify the expressions (\ref{tQ}), (\ref{quarter}):
\begin{eqnarray}\label{utQ}
2cos(\pi P)\mathbf{t}_s(\lambda)=\mathbf{Q}_{+}(q^{s+\frac{1}{4}}\lambda)\mathbf{Q}_{-}(q^{-s-\frac{1}{4}}\lambda)-\\
(-1)^{4s+1}\mathbf{Q}_{+}(q^{-s-\frac{1}{4}}\lambda)\mathbf{Q}_{-}(q^{s+\frac{1}{4}}\lambda)\nonumber,
\end{eqnarray}
where $s\in\mathbb{Z}/4$, $s\ge 0$. Using the qsW relation it is not hard to obtain the expressions for $\mathbf{t}_s(\lambda)$ operators only in terms of 
$\mathbf{Q}_{+}(\lambda)$ or $\mathbf{Q}_{-}(\lambda)$:
\begin{eqnarray}
\mathbf{t}_s(\lambda)=\mathbf{Q}_{\pm}(q^{s+\frac{1}{4}}\lambda)\mathbf{Q}_{\pm}(q^{-s-\frac{1}{4}}\lambda)
\sum^{s}_{k=-s}
\frac{(-1)^{2(k\pm s)}}{\mathbf{Q}_{\pm}(q^{k+\frac{1}{4}}\lambda)\mathbf{Q}_{\pm}(q^{k-\frac{1}{4}}\lambda)}
\end{eqnarray}
For example for $\mathbf{t}_{\frac{1}{2}}(\lambda)$ we obtain the ``triple
relation'':
\begin{eqnarray}
\mathbf{t}_{\frac{1}{2}}(\lambda)=\frac{\mathbf{Q}_{\pm}(q^{-\frac{3}{4}}\lambda)}{\mathbf{Q}_{\pm}(q^{\frac{1}{4}}\lambda)}-\frac{\mathbf{Q}_{\pm}(q^{\frac{3}{4}}\lambda)\mathbf{Q}_{\pm}(q^{-\frac{3}{4}}\lambda)
}{\mathbf{Q}_{\pm}(q^{\frac{1}{4}}\lambda)\mathbf{Q}_{\pm}(q^{-\frac{1}{4}}\lambda)}+\frac{\mathbf{Q}_{\pm}(q^{\frac{3}{4}}\lambda)}{\mathbf{Q}_{\pm}(q^{-\frac{1}{4}}\lambda)}
\end{eqnarray}
Due to the unifying relation (\ref{utQ}) and using again (\ref{sW}) we find the fusion relations:
\begin{eqnarray}\label{fusion}
\mathbf{t}_{j}(q^{\frac{1}{4}}\lambda)\mathbf{t}_{j}(q^{-\frac{1}{4}}\lambda)=
\mathbf{t}_{j+\frac{1}{4}}(\lambda)\mathbf{t}_{j-\frac{1}{4}}(\lambda)+(-1)^{4j}, \quad j=0,\frac{1}{4},\frac{1}{2},1, ...   
\end{eqnarray} 
However, as we already mentioned in the introduction these relations are different from fusion relations obtained for (super)algebras $A_1^{(1)}$, $A_2^{(2)}$, $A_2^{(1)}$, $B(0,1)^{(1)}$ (see \cite{1},\cite{4},\cite{5},\cite{plb1}) 
in the sense that in all these cases operators present in the fusion relation
correspond to the representations of the associated algebras. In our 
case we see that artificially constructed auxiliary quarter operators 
do not correspond to any representation of $C_q(2)^{(2)}$.\\ 
\hspace*{5mm}Let's write the fusion relations for $s=\frac{1}{4}$ with 
$\lambda$ multiplied on $q^{\frac{1}{4}}$:
\begin{eqnarray}
\mathbf{t}_{\frac{1}{2}}(q^{\frac{1}{4}}\lambda)=
\mathbf{t}_{\frac{1}{4}}(q^{\frac{1}{2}}\lambda)
\mathbf{t}_{\frac{1}{4}}
(\lambda)+1.
\end{eqnarray} 
Multiplying both sides on $\mathbf{Q}_{\pm}(\lambda)$ we obtain second 
relation of Baxter type:
\begin{eqnarray}\label{baxter2}
\mathbf{Q}_{\pm}(q\lambda)\mp\mathbf{t}_{\frac{1}{4}}(q^{\frac{1}{2}}\lambda)
\mathbf{Q}_{\pm}(q^{-\frac{1}{2}}\lambda)= \mathbf{t}_{\frac{1}{2}}(q^{\frac{1}{4}}\lambda)\mathbf{Q}_{\pm}(\lambda).
\end{eqnarray} 
\section{Truncation of the Fusion Relations}
Though the Baxter $\mathbf{Q}$-operator method is very powerful there exists another approach which is useful for rational values of the central charge \cite{1}, it is the SCFT counterpart of Baxter's commuting transfer matrix method. In this section we will show that when q is the root of unity the system of fusion relations becomes finite.
So, let's put $q^N=\pm1$, $N\in\mathbb{Z}$, $N>0$. Then
\begin{eqnarray}
&&2cos(\pi P)\mathbf{t}_{\frac{N}{2}-\frac{1}{2}-s}(\lambda q^{\frac{N}{2}})=\\
&&\mathbf{Q}_{+}(q^{N-\frac{1}{2}-s}\lambda)\mathbf{Q}_{-}(q^{\frac{1}{4}+s}\lambda)-(-1)^{4s+1}\mathbf{Q}_{+}(q^{\frac{1}{4}+s}\lambda)\mathbf{Q}_{-}(q^{N-\frac{1}{2}-s}\lambda)\nonumber.
\end{eqnarray} 
Here $s\in\mathbb{Z}/4$, $s\le\frac{N}{2}-\frac{1}{2}$. Note also that
\begin{eqnarray}\label{mon}
\mathbf{Q}_{\pm}(q^N\lambda)=e^{\pm\pi i NP}\mathbf{Q}_{\pm}(\lambda).
\end{eqnarray} 
With the use of such a transform one obtains:
\begin{eqnarray}
&&2cos(\pi P)\mathbf{t}_{\frac{N}{2}-\frac{1}{2}-s}(\lambda q^{\frac{N}{2}})=\\
&&e^{\pi iNP}\mathbf{Q}_{+}(q^{-\frac{1}{4}-s}\lambda)\mathbf{Q}_{-}(q^{\frac{1}{4}+s}\lambda)-(-1)^{4s+1}e^{-\pi iNP}\mathbf{Q}_{+}(q^{\frac{1}{4}+s}\lambda)\mathbf{Q}_{-}(q^{-\frac{1}{4}-s}\lambda)\nonumber.
\end{eqnarray}
Recalling the formula (\ref{tQ}) for $\mathbf{t}_{s}$ we arrive to the following expression:
\begin{eqnarray}
&&\mathbf{t}_{\frac{N}{2}-\frac{1}{2}-s}(\lambda q^{\frac{N}{2}})+(-1)^{4s+1}
e^{\pi iNP}\mathbf{t}_s(\lambda)=\\
&&(-1)^{4s+1}i\frac{sin(\pi NP)}{cos(\pi P)}
\mathbf{Q}_{+}(q^{\frac{1}{4}+s}\lambda)
\mathbf{Q}_{-}(q^{-\frac{1}{4}-s}\lambda), \quad s=0, \frac{1}{4},\frac{1}{2},...,
\frac{N}{2}-\frac{1}{2}.
\nonumber
\end{eqnarray}
 For example for $s=0$ we obtain:
\begin{eqnarray}\label{N/2-1/2}
\mathbf{t}_{\frac{N}{2}-\frac{1}{2}}(\lambda q^{\frac{N}{2}})-
e^{\pi iNP}=
-i\frac{sin(\pi NP)}{cos(\pi P)}
\mathbf{Q}_{+}(q^{\frac{1}{4}}\lambda)
\mathbf{Q}_{-}(q^{-\frac{1}{4}}\lambda).
\end{eqnarray}
Using the basic formula (\ref{tQ}) and property (\ref{mon})
$\mathbf{t}_{\frac{N}{2}-\frac{1}{4}}(\lambda)$ and
for $\mathbf{t}_{\frac{N}{2}}(\lambda)$ 
one arrives to the following relations:
\begin{eqnarray}\label{N/2-1/4}
\mathbf{t}_{\frac{N}{2}-\frac{1}{4}}(\lambda)=
i\frac{sin(\pi NP)}{cos(\pi P)}
\mathbf{Q}_{+}(q^{\frac{N}{2}}\lambda)
\mathbf{Q}_{-}(q^{\frac{N}{2}}\lambda),
\end{eqnarray}
\begin{eqnarray}\label{N/2}
\mathbf{t}_{\frac{N}{2}}(\lambda q^{\frac{N}{2}})-
e^{-\pi iNP}=i\frac{sin(\pi NP)}{cos(\pi P)}
\mathbf{Q}_{+}(q^{\frac{1}{4}}\lambda)
\mathbf{Q}_{-}(q^{-\frac{1}{4}}\lambda).
\end{eqnarray}
Comparing (\ref{N/2}) and (\ref{N/2-1/2}) 
we obtain the following simple formula:
\begin{eqnarray}\label{trunc}
\mathbf{t}_{\frac{N}{2}}(\lambda)+\mathbf{t}_{\frac{N}{2}-
\frac{1}{2}}(\lambda)=2cos(\pi NP).
\end{eqnarray}
In the case when $p=\frac{l+1}{N}$, where $l\ge 0$, $l\in \mathbb{Z}$ there 
exist additional number of truncations:
\begin{eqnarray}
\mathbf{t}_{\frac{N}{2}-\frac{1}{4}}(\lambda)=0, 
\quad \mathbf{t}_{\frac{N}{2}}(\lambda)=
\mathbf{t}_{\frac{N}{2}-\frac{1}{2}}(\lambda)=(-1)^{l+1},\\
\mathbf{t}_{\frac{N}{2}-\frac{1}{2}-s}(\lambda q^{\frac{N}{2}})=
(-1)^{4s}\mathbf{t}_{s}(\lambda)(-1)^{l+1}.\nonumber
\end{eqnarray}
The relation (\ref{trunc}) allows us to to rewrite the fusion relation system
in the Thermodynamic Bethe Ansatz Equations \cite{zamolod} of $D_{N'}$ type as in $A_1^{(1)}$ 
case \cite{bazhanov}:
\begin{eqnarray}\label{tba}
&&Y_{s}(\theta+\frac{i\pi\xi}{2})Y_{s}(\theta-\frac{i\pi\xi}{2})=
(1+Y_{s+\frac{1}{2}}(\theta))(1+Y_{s-\frac{1}{2}}(\theta)),\\ 
&&s=\frac{1}{2}, 1, ...,\frac{N'}{2}-\frac{3}{2},\nonumber\\ 
&&Y_{\frac{N'}{2}-1}(\theta+\frac{i\pi\xi}{2})Y_{\frac{N'}{2}-1}(\theta-\frac{i\pi\xi}{2})=(1+Y_{\frac{N'}{2}-\frac{3}{2}}(\theta))\nonumber\\
&&(1+e^{\pi i P\frac{N'}{2}}
\bar{Y}(\theta))(1+e^{-\pi i P\frac{N'}{2}}\bar{Y}(\theta)),\nonumber\\
&&\bar{Y}(\theta+\frac{i\pi\xi}{2})\bar{Y}(\theta-\frac{i\pi\xi}{2})=
1+Y_{\frac{N'}{2}-1}(\theta).\nonumber
\end{eqnarray}
with the use of identification:
\begin{eqnarray} 
&&Y_{2s}(\theta)=t_{s+\frac{1}{4}}(\lambda)t_{s-\frac{1}{4}}(\lambda)(-1)^{4s},\quad\bar{Y}(\theta)=-t_{\frac{N}{2}-\frac{1}{2}}(\lambda),\\
&&\quad\frac{\beta^2}{2}=\frac{2\xi}{1+2\xi},\quad
\lambda=e^{\frac{\theta}{1+2\xi}},\quad N'=2N,\nonumber
\end{eqnarray}
where we have denoted $t_s(\lambda)$ the eigenvalue of $\mathbf{t}_s(\lambda)$.
\section{Discussion}
In this paper we studied algebraic relations arising from integrable 
structure of CFT provided by the SUSY $N$=1 KdV hierarchy. 
As it was already mentioned in \cite{2}, \cite{3}, \cite{5} the construction of the $\mathbf{Q}$-operator as a ``transfer''-matrix corresponding to the infinite-dimensional q-oscillator representation could be also applied to the 
lattice models. The relations like Baxter's and fusion ones will be also valid
in the lattice case because they depend only on the decomposition properties 
of the representations.\\
\hspace*{5mm}The use of supersymmetry in the considered model is that the limit of 
$\log(\mathbf{t}_{\frac{1}{2}}^{(cl)})(\lambda)$ when $\lambda\to\infty$
is very simple due to the expansion \cite{plb2}:
\begin{eqnarray}
\log(\mathbf{t}_{\frac{1}{2}}^{(cl)}(\lambda))=
-\sum^{\infty}_{n=1}c_n 
I^{(cl)}_{2n-1}{\lambda}^{-4n+2},\qquad{\lambda\to \infty}
\end{eqnarray}
where $c_1=\frac{1}{2}$, $c_n=\frac{(2n-3)!!}{2^n n!}$ for $n>1$ and $I^{(cl)}_{2n-1}$ are the integrals of motion of the SUSY $N$=1 KdV. Thus  
$\lim_{\lambda\to \infty}\log(\mathbf{t}_{\frac{1}{2}}^{(cl)}(\lambda))=0$. We hope that such a nice result remains valid in the quantum case. Maybe this 
conjecture will give us possibility to simplify the study of the analytic 
properties of the eigenvalues of $\mathbf{t}$, $\mathbf{Q}$ operators and 
therefore the corresponding integral equations of Destri-de Vega type \cite{2}, \cite{5}.
We will consider 
these questions in the third part of our manuscript \cite{prep}.
 It will be interesting to consider the obtained TBA equations (\ref{tba})
in the context of the perturbed minimal models as it was done in 
\cite{bazhanov}.\\ 
\hspace*{5mm} In the following we also plan to study the quantization of 
$N>1$ SUSY KdV hierarchies, related with super-W conformal/topological 
integrable field theories.

\section*{Acknowledgements}
We are grateful to F.A. Smirnov for useful discussions.
The work was supported by the Dynasty Foundation (A.M.Z.) and 
RFBR grant 03-01-00593 (P.P.K.).

\section*{Appendix 1}
The superalgebra $C(2)$ has six Chevalley generators: 
$h_1$, $h_2$, ${e_1}^{\pm}$, ${e_2}^{\pm}$ (${e_1}^{\pm}$, ${e_2}^{\pm}$ are 
odd) with the following commutation relations:
\begin{eqnarray} 
&&[h_1,h_2]=0,\quad [h_1,{e_2}^{\pm}]=\pm{e_2}^{\pm},
\quad [h_2,{e_1}^{\pm}]=\pm{e_1}^{\pm},\\
&&ad^{2}_{{e_1}^{\pm}}{e_2}^{\pm}=0, \quad ad^{2}_{{e_2}^{\pm}}{e_1}^{\pm}=0, 
\nonumber\\
&&[h_{\alpha},{e_{\alpha}}^{\pm}]=0 \quad(\alpha =1,2),\quad
[{e_{\beta}}^{\pm},{e_{\beta'}}^{\mp}]=\delta_{\beta, \beta'}h_{\beta}\quad
(\beta , \beta' =1,2),\nonumber
\end{eqnarray}
The fundamental 3-dimensional representation is:
\begin{eqnarray}
&&e_{1}^{+}=
\left(\begin{array}{ccc}
0\quad & 1 \quad&  0\\
0 \quad&  0 \quad&  0\\
0 \quad & 0\quad & 0\\
\end{array}\right),\quad
e_{1}^{-}=
\left(\begin{array}{ccc}
0 \quad & 0 \quad& 0\\
1\quad & 0\quad & 0\\
0 \quad & 0 \quad& 0\\
\end{array}\right),\quad
h_{1}=
\left(\begin{array}{ccc}
1 \quad & 0\quad & 0\\
0 \quad & 1\quad & 0\\
0 \quad & 0\quad & 0\\
\end{array}\right),
\\
&&e_{2}^{+}=
\left(\begin{array}{ccc}
0 \quad & 0\quad & 0\\
0\quad & 0\quad & 1\\
0 \quad & 0\quad & 0\\
\end{array}\right), \quad
e_{2}^{-}=
\left(\begin{array}{ccc}
0\quad & 0\quad & 0\\
0 \quad & 0 \quad & 0\\
0 \quad & -1 \quad & 0\\
\end{array}\right),\quad
h_{2}=
\left(\begin{array}{ccc}
0\quad & 0\quad & 0\\
0\quad & -1\quad & 0\\
0\quad & 0\quad & -1\\
\end{array}\right)\nonumber.
\end{eqnarray}
The following correspondence:
\begin{eqnarray}
&&e_{\alpha_1}\to \lambda(e_1^+ + e_2^+), \quad
e_{-\alpha_1}\to \lambda(e_1^- + e_2^-), \quad 
h_{\alpha_1}\to h_1+h_2,\\
&&e_{\alpha_0}\to \lambda(e_1^- -e_2^-), \quad
e_{-\alpha_0}\to \lambda(-e_1^+ + e_2^+), \quad 
h_{\alpha_0}\to -h_1-h_2\nonumber
\end{eqnarray}
give the evaluation representation (we will consider representations only 
of this type) for the $C(2)^{(2)}$ twisted Kac-Moody 
superalgebra with Chevalley generators $h_{\alpha_{0,1}}$, $e_{{\pm\alpha_1}}$, $e_{\pm\alpha_0}$:
\begin{eqnarray}
&&[h_{\alpha_1},h_{\alpha_0}]=0,\quad [h_{\alpha_0},e_{\pm\alpha_1}]= 
\mp e_{\pm\alpha_1},\quad [h_{\alpha_1},e_{\pm\alpha_0}]= 
\mp e_{\pm\alpha_0},\\
&&[h_{\alpha_i},e_{\pm\alpha_i}]=\pm e_{\pm\alpha_i},\quad
[e_{\pm\alpha_i}, e_{\mp\alpha_j}]=\delta_{i,j}h_{\alpha_i},\quad (i,j=0,1),
\nonumber\\
&&ad^{3}_{e_{\pm\alpha_0}} e_{\pm\alpha_1}=0, \quad
ad^{3}_{e_{\pm\alpha_1}} e_{\pm\alpha_0}=0\nonumber
\end{eqnarray} 
This superalgebra has a usual quantum generalization $C(2)_q^{(2)}$ 
\cite{tolstkhor}:
\begin{eqnarray}
&&[h_{\alpha_0},h_{\alpha_1}]=0,\quad [h_{\alpha_0},e_{\pm\alpha_1}]= 
\mp e_{\pm\alpha_1},\quad [h_{\alpha_1},e_{\pm\alpha_0}]= 
\mp e_{\pm\alpha_0},\\
&&[h_{\alpha_i},e_{\pm\alpha_i}]=\pm e_{\pm\alpha_i}\quad (i=0,1),\quad
[e_{\pm\alpha_i}, e_{\mp\alpha_j}]=\delta_{i,j}[h_{\alpha_i}]\quad (i,j=0,1),
\nonumber\\
&&[e_{\pm\alpha_1},[e_{\pm\alpha_1},[e_{\pm\alpha_1},
e_{\pm\alpha_0}]_{q}]_{q}]_{q}=0,\quad 
[[[e_{\pm\alpha_1},e_{\pm\alpha_0}]_{q},e_{\pm\alpha_0}]_{q},
e_{\pm\alpha_0}]_{q}=0,
\nonumber
\end{eqnarray}
where $[h]=\frac{q^h-q^{-h}}{q-q^{-1}}$, $p(h_{\alpha_{0,1}})=0$, 
$p(e_{\pm \alpha_{0,1}})=1$ and 
q-supercommutator is defined in the following way:
$[e_{\gamma},e_{\gamma'}]_{q}\equiv e_{\gamma}e_{\gamma'} - 
(-1)^{p(e_{\gamma})p(e_{\gamma'})}
q^{(\gamma,\gamma')}e_{\gamma'}e_{\gamma}$, $q=e^{i\pi\frac{\beta^2}{2}}$.
The corresponding coproducts are:
\begin{eqnarray}
&&\Delta(h_{\alpha_j})=h_{\alpha_j}\otimes 1 + 1\otimes h_{\alpha_j},
\quad
\Delta(e_{\alpha_j})=e_{\alpha_j}\otimes q^{h_{\alpha_j}}+1\otimes 
e_{\alpha_j},\\
&&\Delta(e_{-\alpha_j})=e_{-\alpha_j}
\otimes 1+q^{-h_{\alpha_j}} \otimes e_{-\alpha_j}.\nonumber
\end{eqnarray}
The associated universal R-matrix can be expressed in such a way \cite{tolstkhor}:
\begin{equation}
\mathbf{R}=K\bar{\mathbf{R}}=KR_{+}R_{0}R_{-},
\end{equation} 
where 
\begin{eqnarray}
&&K=q^{h_{\alpha}\otimes h_{\alpha}},\quad 
R_{+}=\prod_{n\ge 0}^{\to}R_{n\delta+\alpha},\quad  
R_{-}=\prod_{n\ge 1}^{\gets}R_{n\delta-\alpha},\\
&&R_{0}=\exp((q-q^{-1})\sum_{n>0}d(n)e_{n\delta}\otimes e_{-n\delta})\nonumber.
\end{eqnarray}
and 
$\bar{\mathbf{R}}$ is usually called ``reduced'' universal R-matrix.\\
Here 
\begin{eqnarray}
&&R_{\gamma}=\exp_{(-q^{-1})}(A(\gamma)(q-q^{-1})(e_{\gamma}
\otimes e_{-\gamma})), \quad d(n)=\frac{n(q-q^{-1})}{q^n-q^{-n}},\\ 
&&A(\gamma)=\{(-1)^n\quad if\quad  \gamma=n\delta+\alpha; (-1)^{n-1} \quad if
\quad \gamma=n\delta-\alpha\}\nonumber. 
\end{eqnarray}
The generators $e_{n\delta}$, $e_{n\delta\pm\alpha}$ are defined via the 
q-commutators of Chevalley generators, for example: 
$e_{\delta}=[e_{\alpha_0}, e_{\alpha_1}]_{q^{-1}}$ and 
$e_{-\delta}=[e_{-\alpha_1},e_{-\alpha_0}]_q$. The elements 
$e_{n\delta\pm\alpha}$ are expressed as multiple commutators of 
$e_{\delta}$ with corresponding Chevalley generators, $e_{n\delta}$ ones 
have more complicated form \cite{tolstkhor}.\\
\hspace*{5mm}
We have found the evaluation representations $\rho_s$ of $C_q(2)^{(2)}$ 
which we have called ``$osp_q(1|2)$-induced'' because
each triple $h_{\alpha_i}$, $e_{\alpha_i}$, $e_{-\alpha_i}$ forms an 
$osp_q(1|2)$ subalgebra and their representations are irreducible in $\rho_s$.
 The representations are $4s+1$-dimensional, where s is an integer or 
half-integer. 
Labeling the vectors in the representation space as $|s,l\rangle$ where 
$l=-s,-s-1/2,..,s$ we can write an explicit formulae for the action of the Chevalley generators on these vectors: 
\begin{eqnarray}\label{repr}
&&h_{\alpha_1}|s,l\rangle=-h_{\alpha_0}|s,l\rangle=2l|s,l\rangle\\
&&\lambda^{-1}e_{\alpha_1}|s,l\rangle=c[s-l][s+l+1/2]^{+}
|s,l+1/2\rangle, \quad s-l\in\mathbb{Z}\nonumber\\
&&\lambda^{-1}e_{\alpha_1}|s,l\rangle=c[s-l]^{+}[s+l+1/2]
|s,l+1/2\rangle, \quad s-l\in\mathbb{Z}+1/2\nonumber\\
&&\lambda e_{-\alpha_1}|s,l\rangle=
|s,l-1/2\rangle,\quad s-l\in\mathbb{Z}\nonumber\\
&&\lambda e_{-\alpha_1}|s,l\rangle=-
|s,l-1/2\rangle,\quad s-l\in\mathbb{Z}+1/2\nonumber\\
&&\lambda^{-1}e_{\alpha_0}|s,l\rangle=
|s,l-1/2\rangle,\quad s-l\in\mathbb{Z}\nonumber\\
&&\lambda^{-1}e_{\alpha_0}|s,l\rangle=
|s,l-1/2\rangle,\quad s-l\in\mathbb{Z}+1/2\nonumber\\
&&\lambda e_{-\alpha_0}|s,l\rangle=c[s-l][s+l+1/2]^{+}
|s,l+1/2\rangle,\quad s-l\in\mathbb{Z}\nonumber\\
&&\lambda e_{-\alpha_0}|s,l\rangle=-c[s-l]^{+}[s+l+1/2]
|s,l+1/2\rangle,\quad s-l\in\mathbb{Z}+1/2,\nonumber
\end{eqnarray}
where $[x]=\frac{q^x-q^{-x}}{q-q^{-1}}$, 
$[x]^{+}=\frac{q^x+q^{-x}}{q+q^{-1}}$, $c=([1/2]^{+})^{-1}$.\\
\hspace*{5mm}We also note (this will be important in taking the supertrace) that we 
choose the highest weight $|s,s\rangle$ to be even.\\
\hspace*{5mm}The representations $\rho_s^{+}$ of 
$C_q(2)^{(2)}$, which play also 
a crucial role in the construction of the $\mathbf{Q}$-operators 
(see Sec. 3) are infinite-dimensional and their explicit form is similar 
to that of (\ref{repr}) but with $l$ going from $s$ to $-\infty$.
From a point of view of the $osp_q(1|2)$ subalgebra, generated by
the triple $\{h_{\alpha_1},e_{\alpha_1},e_{-\alpha_1}\}$ these representation
look like  Verma module for $osp_q(1|2)$ with the highest weight 
$|s,s\rangle.$  
One can easily obtain that $\rho_s$ representations arise from the factor of two infinite-dimensional representations of $\rho^{+}$-type: 
$\rho^{+}_s$/$\rho^{+}_{-s-\frac{1}{2}}$. 
\section*{Appendix 2}
The product of two $\mathbf{A}$-operators (\ref{AA}) can be rewritten in the 
following form:
\begin{eqnarray}\label{AAL}
&&\mathbf{A}_{+}(\lambda\mu)\mathbf{A}_{-}(\lambda\mu^{-1})=\\
&&Z_{+}(P)^{-1}Z_{-}(P)^{-1}str_{\chi_{+}\otimes\chi_{-}}e^{i\pi\mathcal{H}}(\mathbf{L}_{+}(\lambda\mu)\otimes I)(I\otimes\mathbf{L}_{-}(\lambda\mu^{-1}))\nonumber,
\end{eqnarray}
where $\mu=q^{s+\frac{1}{4}}$ and  $\mathcal{H}=H\otimes I-I\otimes H$.\\
\hspace*{5mm}By the fundamental property of the universal R-matrix:
$(\Delta\otimes I)\mathbf{R}=\mathbf{R}^{13}\mathbf{R}^{23}$ to rewrite
the tensor product of two $\mathbf{L}$-operators from the previous formula in such a way:
\begin{eqnarray} 
(\mathbf{L}_{+}(\lambda\mu)\otimes I)(I\otimes\mathbf{L}_{-}(\lambda\mu^{-1}))
=\\
e^{\pi iP\mathcal{H}}Pexp^{(q)}\int^{2\pi}_{0}\d u
(\lambda(W_{-}(u)\bar{e}_{\alpha_1} +W_{+}(u)\bar{e}_{\alpha_0})),\nonumber
\end{eqnarray}
where
\begin{eqnarray}
&&\bar{e}_{\alpha_1}=\mu\varepsilon_+\otimes 
q^{-H}+I\otimes\mu^{-1}\varepsilon_-=\alpha_-+\beta_-,\\ 
&&\bar{e}_{\alpha_1}=-\mu\varepsilon_-\otimes q^{H}+I\otimes\mu^{-1}\varepsilon_+=\alpha_++\beta_+.\nonumber
\end{eqnarray}
$\alpha_{\pm}$, $\beta_{\pm}$ and $\mathcal{H}$ satisfy the following commutation relations:
\begin{eqnarray} 
&&\alpha_{\sigma_1}\beta_{\sigma_2}=
-q^{\sigma_1\sigma_2}\beta_{\sigma_2}\alpha_{\sigma_1}, \quad
[\mathcal{H},\alpha_{\pm}]=\mp\alpha_{\pm}, \quad[\mathcal{H},\beta_{\pm}]=\mp\beta_{\pm},\\
&&q^{1/2}\alpha_-\alpha_++q^{-1/2}\alpha_+\alpha_-=-\frac{\mu^2}{q-q^{-1}},
\quad
q^{1/2}\beta_+\beta_-+q^{-1/2}\beta_-\beta_+=\frac{\mu^{-2}}{q-q^{-1}}\nonumber
\end{eqnarray}
Let's choose as $\chi_{\pm}$ (\ref{chi}) the highest weight representations of the q-oscillator algebra ($\ref{q-osc}$) with highest weight vectors 
$|0\rangle_{\pm}$ such that:
\begin{eqnarray} 
\chi_{\pm}(H)|0\rangle_{\pm}=0, \quad \chi_{\pm}(\varepsilon_{\pm})
|0\rangle_{\pm}=0, \quad\chi_{\pm}(\varepsilon_{\mp}^{k})
|0\rangle_{\pm}=|k\rangle_{\pm}
\end{eqnarray}
The tensor product of these modules can be 
decomposed in the following way:
$\chi_+\otimes\chi_-=\oplus_{m=0}^{\infty}\chi^{(m)}$, where 
$\chi^{(m)}$ is the space spanned by the following vectors:
\begin{eqnarray}
|\chi_k^{(m)}\rangle=(\alpha_++\beta_+)^k (\alpha_+-\eta\beta_+)^m
|0\rangle_{+}\otimes|0\rangle_{-}
\end{eqnarray}
where $\eta$ takes such values when $|\chi_k^{(m)}\rangle$ are linearly 
independent. It is easy to see how $\mathcal{H}$ and $\bar{e}_{\alpha_0}$ act on these vectors:
\begin{eqnarray}
\mathcal{H}|\chi_k^{(m)}\rangle=-(m+k)|\chi_k^{(m)}\rangle, \quad 
\bar{e}_{\alpha_0}|\chi_k^{(m)}\rangle=|\chi_{k+1}^{(m)}\rangle
\end{eqnarray}
The action of $\bar{e}_{\alpha_1}$ is more complicated:
\begin{eqnarray}
\bar{e}_{\alpha_1}|\chi_k^{(m)}\rangle=a(l,s)|\chi_{k-1}^{(m)}\rangle + 
b(l,s)|\chi_{k}^{(m-1)}\rangle
\end{eqnarray}
where
\begin{eqnarray}
&&a(l,s)=c[s-l][s+l+1/2]^{+},\quad  s-l\in\mathbb{Z}\\
&&a(l,s)=c[s-l]^{+}[s+l+1/2], \quad s-l\in\mathbb{Z}+1/2\nonumber
\end{eqnarray}
and $k=2(s-l)$, the coefficients $b(l,s)$ can be calculated but we don't need them in the following. The elements 
$\lambda\bar{e}_{\alpha_{1,0}}$ and $\mathcal{H}$ acting in the space 
spanned by $|\chi_k^{(m)}\rangle$ for any constant $m$ coincide up to a constant shift in $\mathcal{H}$ and shift in the ``$m-1$-direction'' of 
$\lambda\bar{e}_{\alpha_{1,0}}$ action coincides with 
action of the corresponding Chevalley generators in the infinite dimensional $\rho_s^{+}$ evaluation representations (\ref{repr}). 
When we take the trace in (\ref{AAL}) there is no contribution of shift 
in ``$m-1$-direction'' and the constant shift in  $\mathcal{H}$ action can be easily calculated, that is:
\begin{eqnarray}
str_{\chi_{+}\otimes\chi_{-}}e^{i\pi\mathcal{H}}(\mathbf{L}_{+}(\lambda\mu)
\otimes\mathbf{L}_{-}(\lambda\mu^{-1}))=
\sum_{m=0}^{\infty}(-1)^m e^{(-2s-m)2\pi i P}\mathbf{t}^{+}_s(\lambda).
\end{eqnarray}
Now recalling the definition of $Z_{\pm}(P)$ we find that in chosen highest weight representation
\begin{eqnarray}
Z_{\pm}(P)=\frac{e^{\pi iP}}{2cos(\pi P)}.
\end{eqnarray}
Therefore,
\begin{equation}
\mathbf{A}_{+}(q^{s+\frac{1}{4}}\lambda)\mathbf{A}_{-}(q^{-s-\frac{1}{4}}
\lambda)=2cos(\pi P)e^{-4\pi i P(s+\frac{1}{4})}\mathbf{t}^{+}_s(\lambda).
\end{equation}

\end{document}